# Double Detonation of Sub-Chandrasekhar White Dwarfs and Subluminous Type Ia Supernovae

C. Sivaram and Kenath Arun

Indian Institute of Astrophysics, Bangalore

**Abstract:** Type Ia supernovae are thought to result from thermonuclear explosions of carbon-oxygen white dwarf (WD) stars. This model generally explains the observed properties with certain exceptions, like sub-luminous supernovae. Here we discuss the possibility of sub-Chandrasekhar WDs detonating due to the build up of a layer of helium on the C-O WD by accreting from a helium rich companion star to explain observed deviations such as subluminous type Ia. We also detail some of the energetics involved that will make such scenarios plausible.

Type Ia supernovae form the highest luminosity class of supernovae and are consequently used as distance indicators over vast expanses of space-time, i.e., over cosmological scales. They are used commonly as the brightest standard candles as they are thought to result from thermonuclear explosions of carbon-oxygen white dwarf stars [1]. These explosions arise when the C-O WD accretes material from a companion star and is pushed over the Chandrasekhar limit causing it to collapse gravitationally and heat up to temperatures of $\sim 7\times10^{8}$ K when carbon detonates.

The degeneracy (i.e. high density of C and O nuclei) accelerate the reaction rate so that the entire white dwarf can be incinerated and disintegrated resulting in about 0.5 – 1.0 solar mass of Ni-56, which subsequently undergoes two consecutive beta decays (6 days to Co-56 and 77 days to Fe-56), the exponential decay of these isotopes then powering the light curve for a few months releasing at least $10^{42}-10^{43}$ J in the optical band. These models [2] generally explain the observed properties, with notable exceptions like the sub-luminous 1991bg type of SN [3].

It has also been debated whether all progenitors of SN Ia are single white dwarfs pushed over the limit. Mergers of WDs (in a binary, for e.g. white dwarf binaries with 5 minute orbital periods are known) could give rise to SN Ia [4, 5].

However some calculations did not result in an explosion [6]. More recently it was suggested that merger of equal mass WDs could lead to sub-luminous explosions [7]. Again in such sub-luminous explosions, the C-O nuclei would not be expected to be completely converted to mostly Ni-56. For instance isotopes like Ti-50, are supposed to be primarily produced in such so called sub-Chandrasekhar SN Ia [8]. In such collapses electron captures may dominate to produce neutron rich nuclei like Ti-50.

We also have the recent example of SN2005E, which showed presence of about 0.3 solar mass of Calcium (most Calcium rich SN), which is an intermediate stage in the production of Ni. So this is an example of incomplete silicon burning occurring in low density C-O fuel for a range of temperatures. The density of a WD scales as the mass, M squared, i.e. $\rho \propto M^2$.

A very large number of white dwarfs are known to have a mass substantially lower than a solar mass [9]. Is there any way these sub-Chandrasekhar WDs could detonate?

One way could be to build up a layer of helium on the C-O WD by accreting from a helium rich companion star, i.e. a H deficient star with an extensive He atmosphere. The helium layer would first detonate at $T \sim 10^8$K releasing enough energy to heat the C-O nuclei to $7 \times 10^8$ K to initiate C-burning, which would incinerate a sub-Chandrasekhar WD. The lower progenitor mass would then give rise to a subluminous SN type Ia.

Here we detail some energetic of the phenomena which would make such scenarios plausible. Take a 0.6 solar mass WD as an example. Its gravitational binding energy (GBE) would be:

$$E_G \approx \frac{GM^2}{R} \approx 7 \times 10^{49} \text{ ergs} \qquad \dots (1)$$

As one gram of carbon, undergoing nuclear detonation releases $\sim 7 \times 10^{17}$ ergs, $10^{32}$ g of carbon must detonate to form Ni-56 to disintegrate the white dwarf. This mass is ~0.05 solar mass. To detonate carbon, the temperature required is $\sim 7 \times 10^9$ K. The required energy to heat the WD to this temperature is:

$$W \approx MR_gT \approx 7 \times 10^{49} \text{ ergs} \qquad \dots (2)$$

(where $R_g \sim 10^8$ ergs/g is the gas constant)

The helium layer which forms on the WD must be heated to a temperature of $\sim 10^8$K to trigger helium burning. The height of the layer (on the WD) required to heat the helium to this temperature ($T_{He}$) is:

$$h \approx \frac{R_g T_{He}}{g_{WD}} \quad \text{... (3)}$$

Where, $g_{WD} \approx 3\times10^8$ cm/s$^2$ is the acceleration due to gravity on the WD. Thus:

$$h \approx \frac{10^8 \times 10^8}{3\times10^8} \sim 3\times10^7 \text{ cm} \quad \text{... (4)}$$

The mass of the He layer is:

$$M_l = 4\pi R^2 h\rho \approx 2\times10^{31}\,\text{g} \quad \text{... (5)}$$

(where, $\rho \sim 10^5\, g/cc,\ R_{WD} \sim 10^9\, cm$) which is ~0.01 solar mass

As the helium nuclear reaction releases $3\times10^{18}$ ergs/g, the detonation of the helium layer (on reaching the reaction temperature of $\sim 10^8$K) would release an energy of $\sim 6\times10^{49}$ ergs, which is sufficient to heat the C-O WD, to the required temperature for carbon burning. So this double detonation mechanism, first of the helium layer accumulating on the WD, and followed by detonation of the sub-Chandrasekhar C-O WD, could result in a sub-luminous type Ia SN.

A lower mass He layer could detonate an even lower mass C-O WD, and a heavier mass He-shell can detonate a heavier WD. For 1.3 solar mass WD, we need a 0.1 solar mass layer, since the GBE scales as $M^{7/3}$.

The collapse time scale for the WD is about a second but as the reactions have much shorter time scales, the explosions of the WD is inevitable. Our analytical results are in agreement with numerical calculations of other authors. [10]

One can also consider a situation when a white dwarf close to the Chandrasekhar limit acquires such a helium layer or debris from an accretion disc (or tidal disruption of a low mass object), falls onto the WD [11]. The WD, in this case, which may be 'Super Chandrasekhar', would collapse, but the temperature to which it would be heated up (as the energy released scales as $M^{7/3}$) would be substantially higher than the required $7\times10^8$K for carbon detonation. There would also be now losses due to the photoneutrino process which scales at least as $T^8$. So even if the WD mass is ten percent higher than the limit, the neutrino energy loss would increase by a factor of three or more ($T$ would go up by $M^{4/3}$, so the loss rate would increase as $M^{11}$) [12]. So rather than the detonated disintegration of the WD, we would have the collapse of the WD, followed by e-capture by the heavier nuclei, leading to a neutron star.

Moreover, there is a general relativistic induced instability in the collapse of WD's (above the mass limit). This sets in at about 250 times the Schwarzschild radius [13] (i.e. at <1000km). This would inevitably lead to collapse to a NS for a super-Chandrasekhar WD (rather than a nuclear detonation induced fragmentation).

In summary, we have considered the possibility of SN type Ia being produced by sub-Chandrasekhar and super-Chandrasekhar WD's, rather than only the canonical limiting mass white dwarf.